\documentclass[final]{cvpr}

\usepackage{times}
\usepackage{epsfig}
\usepackage{graphicx}
\usepackage{amsmath}
\usepackage{amssymb}
\usepackage[table,x11names]{xcolor}

\usepackage[pagebackref=true,breaklinks=true,colorlinks,bookmarks=false]{hyperref}

\def\cvprPaperID{****} 
\def\confYear{CVPR 2021}
\definecolor{grayhighlight}{RGB}{213,229,255}

\newsavebox\CBox
\def\textBF#1{\sbox\CBox{#1}\resizebox{\wd\CBox}{\ht\CBox}{\textbf{#1}}}

\begin{document}

\title{Fast Camera Image Denoising on Mobile GPUs with Deep Learning,\\ Mobile AI 2021 Challenge: Report}
\author{
Andrey Ignatov \and Kim Byeoung-su \and Radu Timofte \and Angeline Pouget \and
Fenglong Song \and Cheng Li \and Shuai Xiao \and Zhongqian Fu \and Matteo Maggioni \and Yibin Huang \and
Shen Cheng \and  Xin Lu \and Yifeng Zhou \and Liangyu Chen \and Donghao Liu \and Xiangyu Zhang \and Haoqiang Fan \and Jian Sun \and Shuaicheng Liu \and
Minsu Kwon \and Myungje Lee \and Jaeyoon Yoo \and Changbeom Kang \and Shinjo Wang \and
Bin Huang \and Tianbao Zhou \and
Shuai Liu \and Lei Lei \and Chaoyu Feng \and
Liguang Huang \and Zhikun Lei \and Feifei Chen
}

\maketitle

\begin{abstract}

Image denoising is one of the most critical problems in mobile photo processing. While many solutions have been proposed for this task, they are usually working with synthetic data and are too computationally expensive to run on mobile devices. To address this problem, we introduce the first Mobile AI challenge, where the target is to develop an end-to-end deep learning-based image denoising solution that can demonstrate high efficiency on smartphone GPUs. For this, the participants were provided with a novel large-scale dataset consisting of noisy-clean image pairs captured in the wild. The runtime of all models was evaluated on the Samsung Exynos 2100 chipset with a powerful Mali GPU capable of accelerating floating-point and quantized neural networks. The proposed solutions are fully compatible with any mobile GPU and are capable of processing 480p resolution images under 40-80 ms while achieving high fidelity results. A detailed description of all models developed in the challenge is provided in this paper.

\end{abstract}
{\let\thefootnote\relax\footnotetext{%
\hspace{-5mm}$^*$
Andrey Ignatov, Kim Byeoung-su and Radu Timofte are the Mobile AI 2021 challenge organizers \textit{(andrey@vision.ee.ethz.ch, rui.kim@samsung.com, radu.timofte@vision.ee.ethz.ch)}. Angeline Pouget performed data collection. The other authors participated in the challenge. Appendix \ref{sec:apd:team} contains the authors' team names and affiliations. \vspace{2mm} \\ Mobile AI 2021 Workshop website: \\ \url{https://ai-benchmark.com/workshops/mai/2021/}
}}

\section{Introduction}

\begin{figure*}[t!]
\centering
\setlength{\tabcolsep}{1pt}
\resizebox{\linewidth}{!}
{
\includegraphics[width=0.5\linewidth]{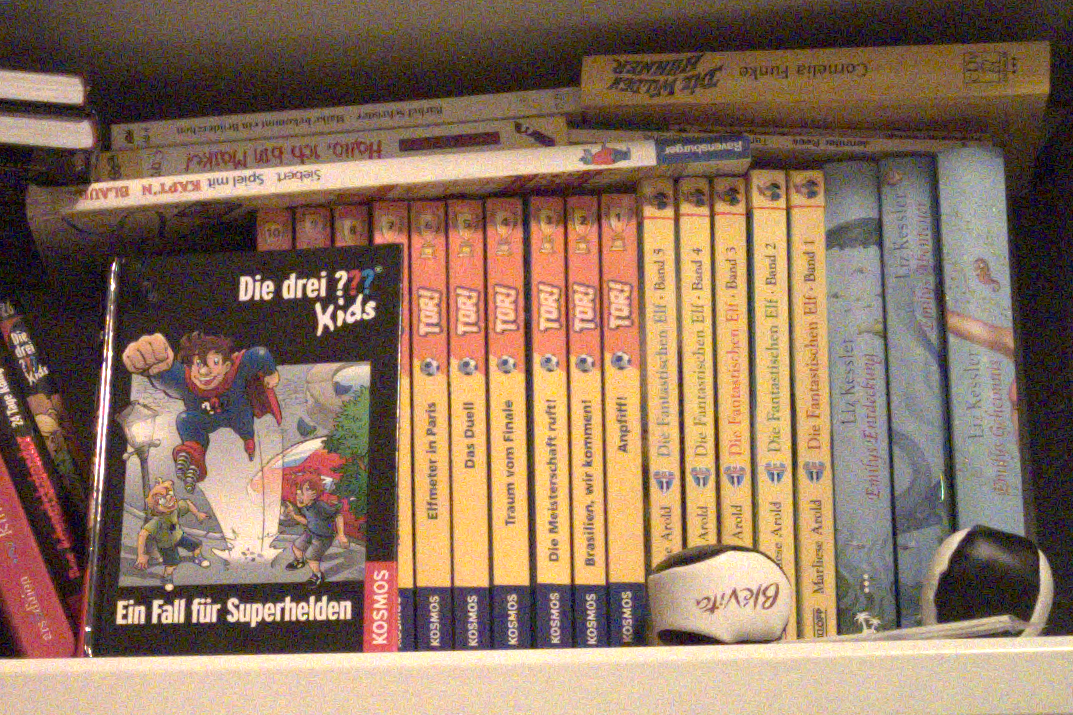} \hspace{2mm}
\includegraphics[width=0.5\linewidth]{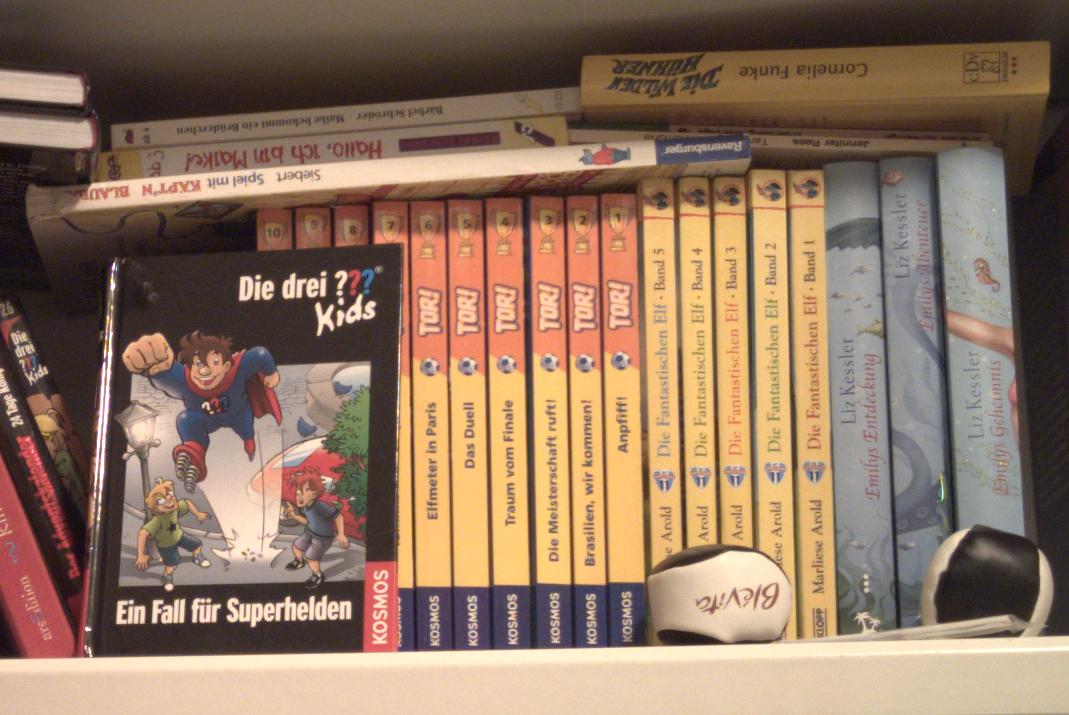}
}
\vspace{0cm}
\caption{Sample crops from the original and denoised images from the collected dataset. Best zoomed on screen.}
\label{fig:example_photos}
\vspace{-0.2cm}
\end{figure*}

Despite the recent advances in mobile camera sensors, image denoising still remains one of the most challenging tasks when it comes to processing mobile photo and video data. The hardware constraints do not allow to significantly increase the size of mobile cameras, which together with increased sensor resolutions and smaller pixels leads to high noise levels on images taken in low-light conditions.
To deal with this problem, many classical approaches have been proposed in the past~\cite{kuan1985adaptive,liu2007automatic,dabov2007image,buades2005non,buades2005review,motwani2004survey,mihcak1999low,starck2002curvelet}. Much better quantitative results were obtained later with CNN-based deep learning approaches~\cite{gu2019brief,tai2017memnet,zhang2018ffdnet,zhang2017beyond,abdelhamed2020ntire,abdelhamed2019ntire}. Despite the good fidelity scores, these works were using either artificial training and validation data~\cite{tai2017memnet,zhang2018ffdnet,zhang2017beyond} or a very small set of indoor images~\cite{abdelhamed2018high,abdelhamed2020ntire,abdelhamed2019ntire}, thus limiting their application to real noisy camera data. Besides that, the proposed methods were not optimized for computational efficiency, which is essential for this and other tasks related to image processing and enhancement~\cite{ignatov2017dslr,ignatov2018wespe,ignatov2020replacing} on mobile devices. In this challenge, we take one step further in solving this problem by using a more advanced real data and by putting additional efficiency-related constraints on the developed solutions.

When it comes to the deployment of AI-based solutions on mobile devices, one needs to take care of the particularities of mobile NPUs and DSPs to design an efficient model. An extensive overview of smartphone AI acceleration hardware and its performance is provided in~\cite{ignatov2019ai,ignatov2018ai}. According to the results reported in these papers, the latest mobile NPUs are already approaching the results of mid-range desktop GPUs released not long ago. However, there are still two major issues that prevent a straightforward deployment of neural networks on mobile devices: a restricted amount of RAM, and a limited and not always efficient support for many common deep learning layers and operators. These two problems make it impossible to process high resolution data with standard NN models, thus requiring a careful adaptation of each architecture to the restrictions of mobile AI hardware. Such optimizations can include network pruning and compression~\cite{chiang2020deploying,ignatov2020rendering,li2019learning,liu2019metapruning,obukhov2020t}, 16-bit / 8-bit~\cite{chiang2020deploying,jain2019trained,jacob2018quantization,yang2019quantization} and low-bit~\cite{cai2020zeroq,uhlich2019mixed,ignatov2020controlling,liu2018bi} quantization, device- or NPU-specific adaptations, platform-aware neural architecture search~\cite{howard2019searching,tan2019mnasnet,wu2019fbnet,wan2020fbnetv2}, \etc.

While many challenges and works targeted at efficient deep learning models have been proposed recently, the evaluation of the obtained solutions is generally performed on desktop CPUs and GPUs, making the developed solutions not practical due to the above mentioned issues. To address this problem, we introduce the first \textit{Mobile AI Workshop and Challenges}, where all deep learning solutions are developed for and evaluated on real mobile devices.
In this competition, the participating teams were provided with a large-scale image denoising dataset obtained with a recent Sony mobile camera sensor capturing photos in the burst mode. The obtained for each scene images were averaged to get a clean photo, and the resulting noisy-clean image pairs were used to train an end-to-end deep learning solution for this task.
Within the challenge, the participants were evaluating the runtime and tuning their models on the Samsung Exynos 2100 platform featuring a powerful Mali-G78 MP14 mobile GPU that can accelerate floating-point and quantized neural networks.
The final score of each submitted solution was based on the runtime and fidelity results, thus balancing between the image reconstruction quality and efficiency of the proposed model. Finally, all developed solutions are fully compatible with the TensorFlow Lite framework~\cite{TensorFlowLite2021}, thus can be deployed and accelerated on any mobile platform providing AI acceleration through the Android Neural Networks API (NNAPI)~\cite{NNAPI2021} or custom TFLite delegates~\cite{TFLiteDelegates2021}.

\smallskip

This challenge is a part of the \textit{MAI 2021 Workshop and Challenges} consisting of the following competitions:

\small

\begin{itemize}
\item Learned Smartphone ISP on Mobile NPUs~\cite{ignatov2021learned}
\item Real Image Denoising on Mobile GPUs
\item Quantized Image Super-Resolution on Mobile NPUs~\cite{ignatov2021real}
\item Real-Time Video Super-Resolution on Mobile GPUs~\cite{romero2021real}
\item Single-Image Depth Estimation on Mobile Devices~\cite{ignatov2021fastDepth}
\item Quantized Camera Scene Detection on Smartphones~\cite{ignatov2021fastSceneDetection}
\item High Dynamic Range Image Processing on Mobile NPUs
\end{itemize}

\normalsize

\noindent The results obtained in the other competitions and the description of the proposed solutions can be found in the corresponding challenge papers.


\begin{figure*}[t!]
\centering
\setlength{\tabcolsep}{1pt}
\resizebox{0.96\linewidth}{!}
{
\includegraphics[width=1.0\linewidth]{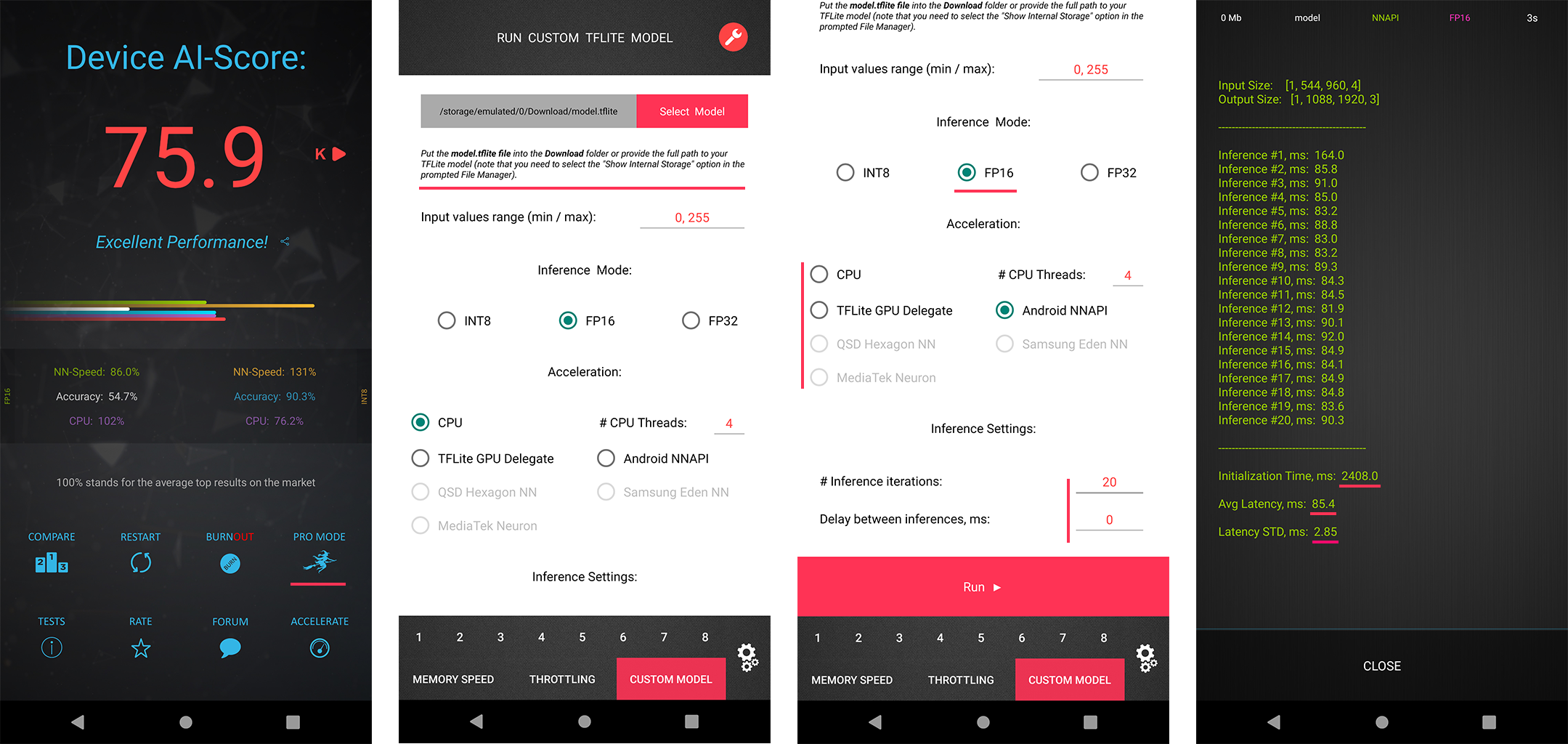}
}
\vspace{0.2cm}
\caption{Loading and running custom TensorFlow Lite models with AI Benchmark application. The currently supported acceleration options include Android NNAPI, TFLite GPU, Hexagon NN, Samsung Eden and MediaTek Neuron delegates as well as CPU inference through TFLite or XNNPACK backends. The latest app version can be downloaded at \url{https://ai-benchmark.com/download}}
\label{fig:ai_benchmark_custom}
\end{figure*}

\section{Challenge}

To develop an efficient and practical solution for mobile-related tasks, one needs the following major components:

\begin{enumerate}
\item A high-quality and large-scale dataset that can be used to train and evaluate the solution on real (not synthetically generated) data;
\item An efficient way to check the runtime and debug the model locally without any constraints;
\item An ability to regularly test the runtime of the designed neural network on the target mobile platform or device.
\end{enumerate}

This challenge addresses all the above issues. Real training data, tools, and runtime evaluation options provided to the challenge participants are described in the next sections.

\subsection{Dataset}

To handle the considered image denoising problem, a large scale dataset consisting of noisy-clean images was collected. For this, we used a recent Sony mobile camera sensor and captured photos in the burst mode: for each scene, 20 images were obtained and then averaged to get a clean photo. The illumination conditions varied from moderate to completely dark throughout the data collection process, the images were shot indoors and outdoors to get a variety of in-the-wild scenes with different noise patterns. The photos of more than 1000 different scenes were obtained and then checked manually to remove out-of-focus images and the ones where the built-in optical image stabilization caused misalignments. It should be additionally mentioned that we shot RAW photos using Android Camera2API~\cite{Camera2API2021} to avoid any effects from smartphones' integrated image denoising ISP modules, and the resulting images were then converted to RGB format using a stand-alone classical ISP system with disabled noise correction options. The resolution of the images was 4000$\times$3000 pixels, around 650 scenes were used for training the models, while the remaining photos were reserved for validation and testing. An example set of collected images is shown in Fig.~\ref{fig:example_photos}.

\begin{table*}[t!]
\centering
\resizebox{\linewidth}{!}
{
\begin{tabular}{l|c|cc|cc|c|c}
\hline
Team \, & \, Author \, & \, Framework \, & \, Model Size, KB \, & \, PSNR$\uparrow$ \, & \, SSIM$\uparrow$ & \, Runtime, ms $\downarrow$ \, & \, Final Score \\
\hline
\hline
NOAHTCV & noahtcv & TensorFlow & 209 & 37.52 & 0.9150 & 39 & \textBF{53.99} \\
Megvii & chengshen & TensorFlow & 14276 & 37.83 & 0.9072 & 84 & 38.52 \\
ENERZAi Research \, & \, Minsu.Kwon \, & TensorFlow & 81 & 36.33 & 0.8930 & \textBF{11} & 36.77 \\
MOMA-Denoise & npzl & TensorFlow & 1404 & 37.37 & 0.9087 & 54 & 31.67 \\
\textit{ENERZAi Research $^*$} & \textit{myungje.lee} & TensorFlow & 118 & 36.22 & 0.9023 & 23 & 15.1 \\
Mier & q935970314 & \, PyTorch / TensorFlow \, & 1528 & 36.34 & 0.9066 & 314 & 1.31 \\
GdAlgo & TuningMan & PyTorch / TensorFlow & 18288 & \textBF{37.84} & \textBF{0.9157} & 5019 & 0.65 \\
\end{tabular}
}
\vspace{2.6mm}
\caption{\small{Mobile AI 2021 Real Image Denoising challenge results and final rankings. The runtime values were obtained on 480p (720$\times$480) images. Team \textit{NOAHTCV} is the challenge winner. $^*$~The second solution from \textit{ENERZAi Research} team did not participate in the official test phase, its scores are shown for general information only.}}
\label{tab:results}
\end{table*}

\subsection{Local Runtime Evaluation}

When developing AI solutions for mobile devices, it is vital to be able to test the designed models and debug all emerging issues locally on available devices. For this, the participants were provided with the \textit{AI Benchmark} application~\cite{ignatov2018ai,ignatov2019ai} that allows to load any custom TensorFlow Lite model and run it on any Android device with all supported acceleration options. This tool contains the latest versions of \textit{Android NNAPI, TFLite GPU, Hexagon NN, Samsung Eden} and \textit{MediaTek Neuron} delegates, therefore supporting all current mobile platforms and providing the users with the ability to execute neural networks on smartphone NPUs, APUs, DSPs, GPUs and CPUs.

\smallskip

To load and run a custom TensorFlow Lite model, one needs to follow the next steps:

\begin{enumerate}
\setlength\itemsep{0mm}
\item Download AI Benchmark from the official website\footnote{\url{https://ai-benchmark.com/download}} or from the Google Play\footnote{\url{https://play.google.com/store/apps/details?id=org.benchmark.demo}} and run its standard tests.
\item After the end of the tests, enter the \textit{PRO Mode} and select the \textit{Custom Model} tab there.
\item Rename the exported TFLite model to \textit{model.tflite} and put it into the \textit{Download} folder of the device.
\item Select mode type \textit{(INT8, FP16, or FP32)}, the desired acceleration/inference options and run the model.
\end{enumerate}

\noindent These steps are also illustrated in Fig.~\ref{fig:ai_benchmark_custom}.

\subsection{Runtime Evaluation on the Target Platform}

In this challenge, we use the \textit{Samsung Exynos 2100} SoC as our target runtime evaluation platform. This chipset contains a powerful 14-core \textit{Mali-G78 GPU} capable of accelerating floating point and quantized models, being ranked among the top three mobile platforms by AI Benchmark at the time of its release~\cite{AIBenchmark202104}. Within the challenge, the participants were able to upload their TFLite models to an external server and get a feedback regarding the speed of their model: the runtime of their solution on the above mentioned Mali GPU or an error log if the model contains some incompatible operations. The models were parsed and accelerated using Samsung Eden delegate designed and tuned for high-end Exynos mobile platforms. The same setup was also used for the final runtime evaluation. The participants were additionally provided with a detailed model optimization guideline demonstrating the restrictions and the most efficient setups for each supported TFLite op.

\subsection{Challenge Phases}

The challenge consisted of the following phases:

\vspace{-0.8mm}
\begin{enumerate}
\item[I.] \textit{Development:} the participants get access to the data and AI Benchmark app, and are able to train the models and evaluate their runtime locally;
\item[II.] \textit{Validation:} the participants can upload their models to the remote server to check the fidelity scores on the validation dataset, to get the runtime on the target platform, and to compare their results on the validation leaderboard;
\item[III.] \textit{Testing:} the participants submit their final results, codes, TensorFlow Lite models, and factsheets.
\end{enumerate}
\vspace{-0.8mm}

\subsection{Scoring System}

All solutions were evaluated using the following metrics:

\vspace{-0.8mm}
\begin{itemize}
\setlength\itemsep{-0.2mm}
\item Peak Signal-to-Noise Ratio (PSNR) measuring fidelity score,
\item Structural Similarity Index Measure (SSIM), a proxy for perceptual score,
\item The runtime on the target Exynos 2100 platform.
\end{itemize}
\vspace{-0.8mm}

The score of each final submission was evaluated based on the next formula ($C$ is a constant normalization factor):

\smallskip
\begin{equation*}
\text{Final Score} \,=\, \frac{2^{2 \cdot \text{PSNR}}}{C \cdot \text{runtime}},
\end{equation*}
\smallskip

During the final challenge phase, the participants did not have access to the test dataset. Instead, they had to submit their final TensorFlow Lite models that were subsequently used by the challenge organizers to check both the runtime and the fidelity results of each submission under identical conditions. This approach solved all the issues related to model overfitting, reproducibility of the results, and consistency of the obtained runtime/accuracy values.

\section{Challenge Results}

From above 190 registered participants, 8 teams entered the final phase and submitted valid results, TFLite models, codes, executables and factsheets. Table~\ref{tab:results} summarizes the final challenge results and reports PSNR, SSIM and runtime numbers for the top solutions on the final test dataset and on the target evaluation platform. The proposed methods are described in section~\ref{sec:solutions}, and the team members and affiliations are listed in Appendix~\ref{sec:apd:team}.

\subsection{Results and Discussion}

Nearly all submitted solutions demonstrated a very high efficiency: the majority of models are able to process one 480p (720$\times$480 px) image under 60 ms on the target Samsung Exynos 2100 SoC, while the reported runtime results on full-resolution camera images are less than 0.8 seconds for most networks. All proposed solutions were derived from a U-Net~\cite{ronneberger2015u} like architecture which is not surprising: when feature maps are downsampled in its encoder block, the noise is also removed very efficiently, thus this model type suits perfectly for the considered problem. The major differences, however, come from the way in which the participants optimized their solutions for better runtime and fidelity results. The challenge winner, team \textit{NOAHTCV}, used Neural Architecture Search (NAS) to find the optimal model design, the same approach was also utilized by \textit{MOMA-Denoise}. \textit{ENERZAi Research} team based its solution on an efficient knowledge transfer approach consisting of the joint training of two (tiny and large) models sharing the same feature extraction block. Their model achieved the fastest runtime and the smallest size (only 81 kilobytes) in this challenge. Another interesting approach was proposed by \textit{Megvii} that presented a modified decoder block splitting and processing feature maps in two parallel channels to reduce the number of multiply-accumulate operations.

The best fidelity scores and visual results were obtained by team \textit{GdAlgo}~-- this was achieved at the price of using a relatively large multiscale model requiring around five seconds to process one image on Mali GPU. It should be also mentioned that though some of the proposed solutions demonstrated nearly the same speed on desktop CPUs and GPUs, their results on the target Samsung platform differ more than 2-5 times. This explicitly shows that the runtime values obtained on common deep learning hardware are not representative when it comes to model deployment on mobile AI silicon: even solutions that might seem to be very efficient can struggle significantly due to the specific constraints of smartphone AI acceleration hardware and frameworks. This makes deep learning development for mobile devices so challenging, though the results obtained in this competition demonstrate that one can get a very efficient model when taking the above aspects into account.

\section{Challenge Methods}
\label{sec:solutions}

\noindent This section describes solutions submitted by all teams participating in the final stage of the MAI 2021 Real Image Denoising challenge.

\subsection{NOAHTCV}

\begin{figure}[h!]
\centering
\resizebox{1.0\linewidth}{!}
{
\includegraphics[width=1.0\linewidth]{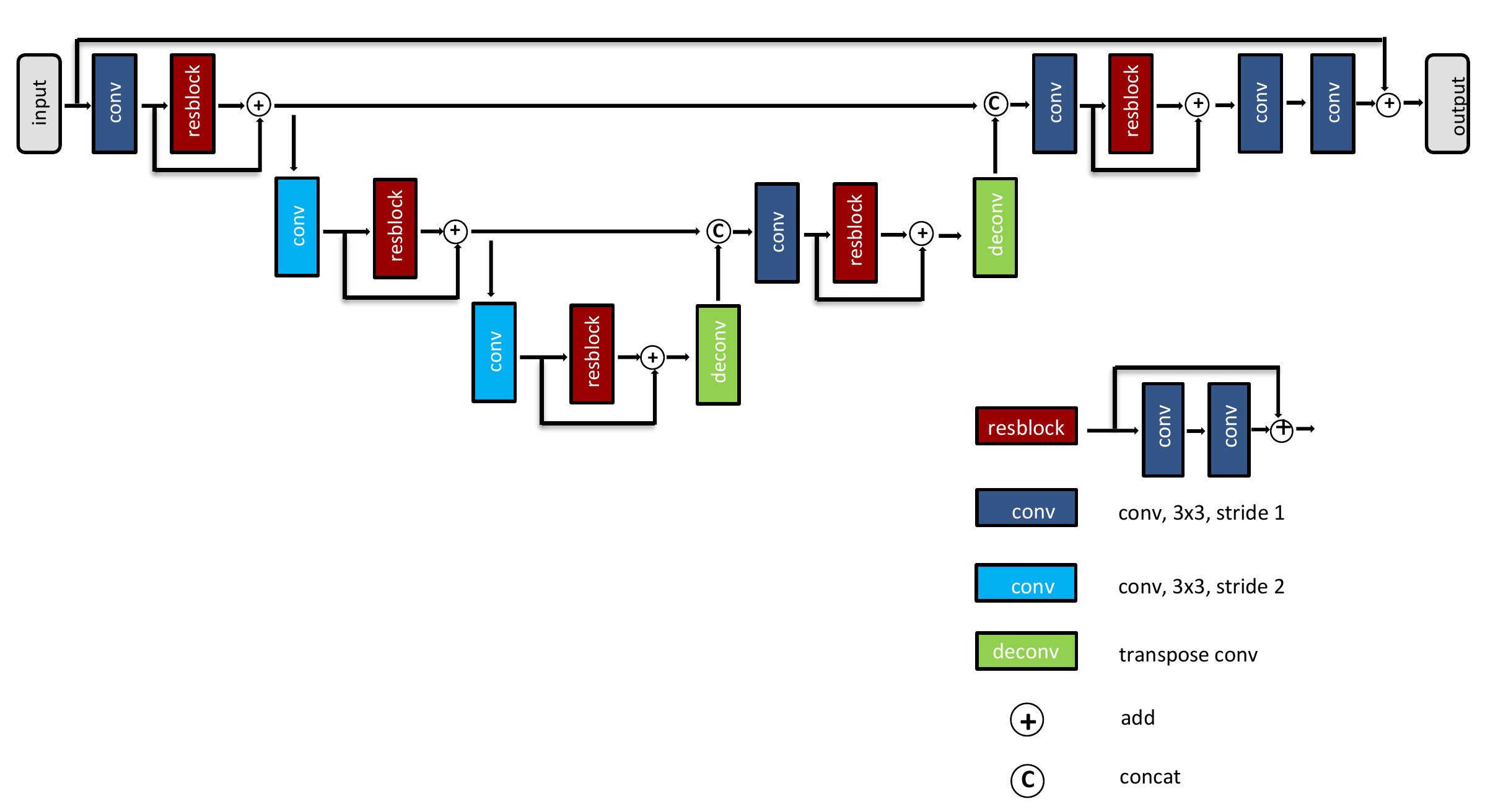}
}
\caption{\small{Image denoising network proposed by team NOAHTCV.}}
\label{fig:NOAHTCV}
\end{figure}

Team NOAHTCV applied Neural Architecture Search (NAS)~\cite{song2020efficient} to find an optimal model for this task. The authors started from a multi-scale architecture and used the challenge scoring formula as a target NAS metric. The space of available operators and layers was narrowed to those that are fully supported on mobile devices. To enhance NAS performance, the authors additionally used knowledge distillation to explore more promising candidates and accelerate the optimization procedure. During the fine-tuning stage, each model candidate was optimized by utilizing both the target clean images and the reconstruction results from a larger pre-trained ``teacher'' model.

Fig.~\ref{fig:NOAHTCV} demonstrates the final model architecture. The authors especially emphasize the role of skip connections on fidelity scores and the effect of upsampling and downsampling operations on the runtime results. The models were trained on patches of size 256$\times$256 pixels using Adam optimizer with a batch size of 64 for 200 epochs. The learning rate was set to $1e-4$ and decreased to $1e-5$ by applying a cosine decay. $L_2$ loss was used as the main fidelity metric.

\subsection{Megvii}

\begin{figure}[h!]
\centering
\resizebox{1.0\linewidth}{!}
{
\includegraphics[width=1.0\linewidth]{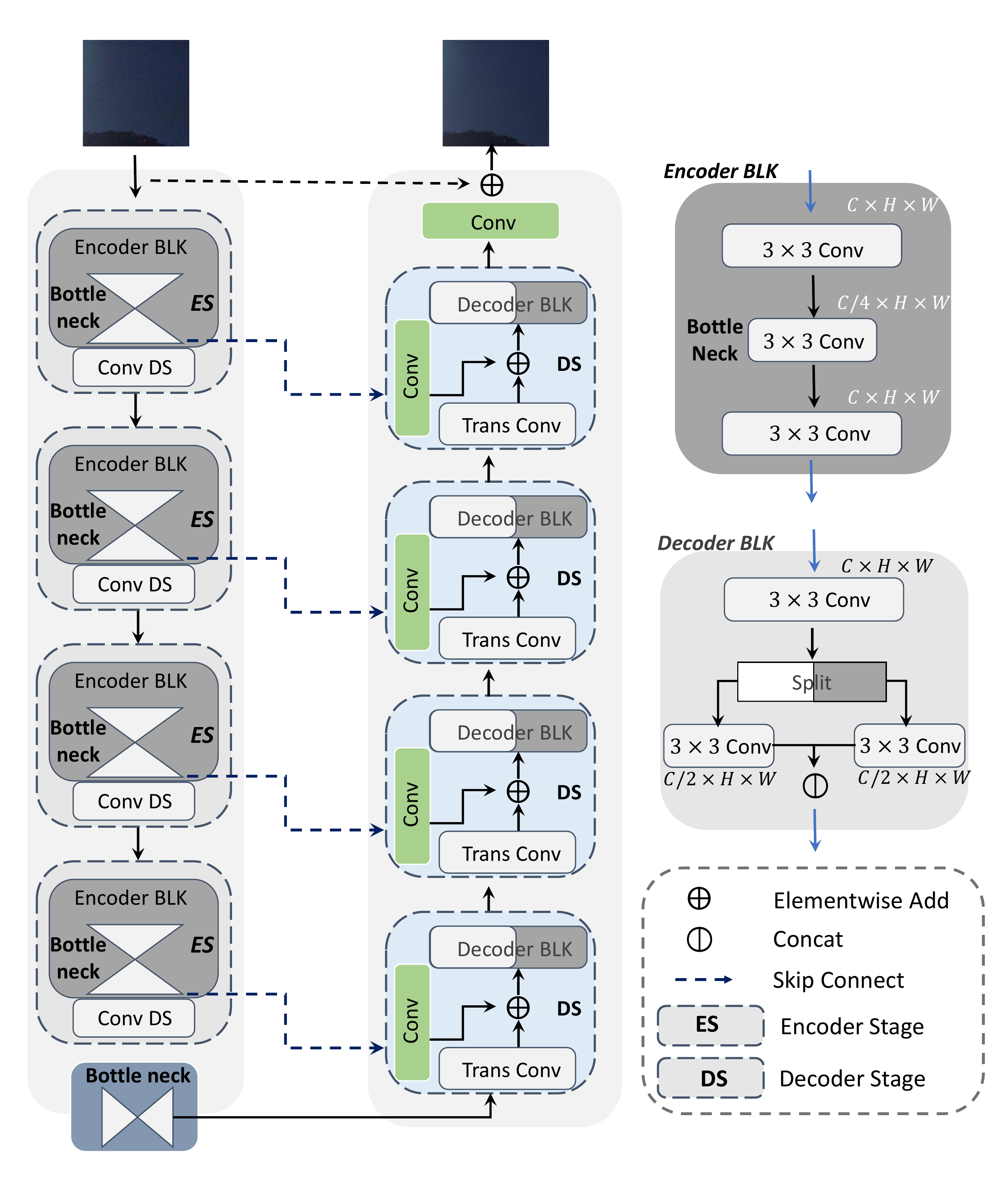}
}
\caption{\small{A U-Net based model with modified decoder blocks presented by team Megvii.}}
\label{fig:Megvii}
\end{figure}

Team Megvii proposed a light U-Net~\cite{ronneberger2015u} based architecture with a modified decoder block (Fig.~\ref{fig:Megvii}). To improve the efficiency of the model, the authors used small 3$\times$3 convolutional filters in all layers, and set the size of the feature maps to be multiple of eight. No residual blocks were used in the model to decrease its runtime. Each decoder module contains two layers: the output of the first convolutional layer is split into two groups and fed to two convolutions of the second layer, both having half of the channels to decrease the number of FLOPs and multiply-accumulate operations (MACs). The model was trained to
maximize PSNR loss using Adam with a batch size of 64 for 162K iterations. The learning rate was set to $2e-4$ and was steadily decreased to $1e-6$ using the cosine annealing strategy. The network was trained on 448$\times$448 patches, vertical and horizontal flips and the Mixup~\cite{zhang2017mixup} strategy were applied for data augmentation.

\subsection{ENERZAi Research}

\begin{figure}[h!]
\centering
\resizebox{1.0\linewidth}{!}
{
\includegraphics[width=1.0\linewidth]{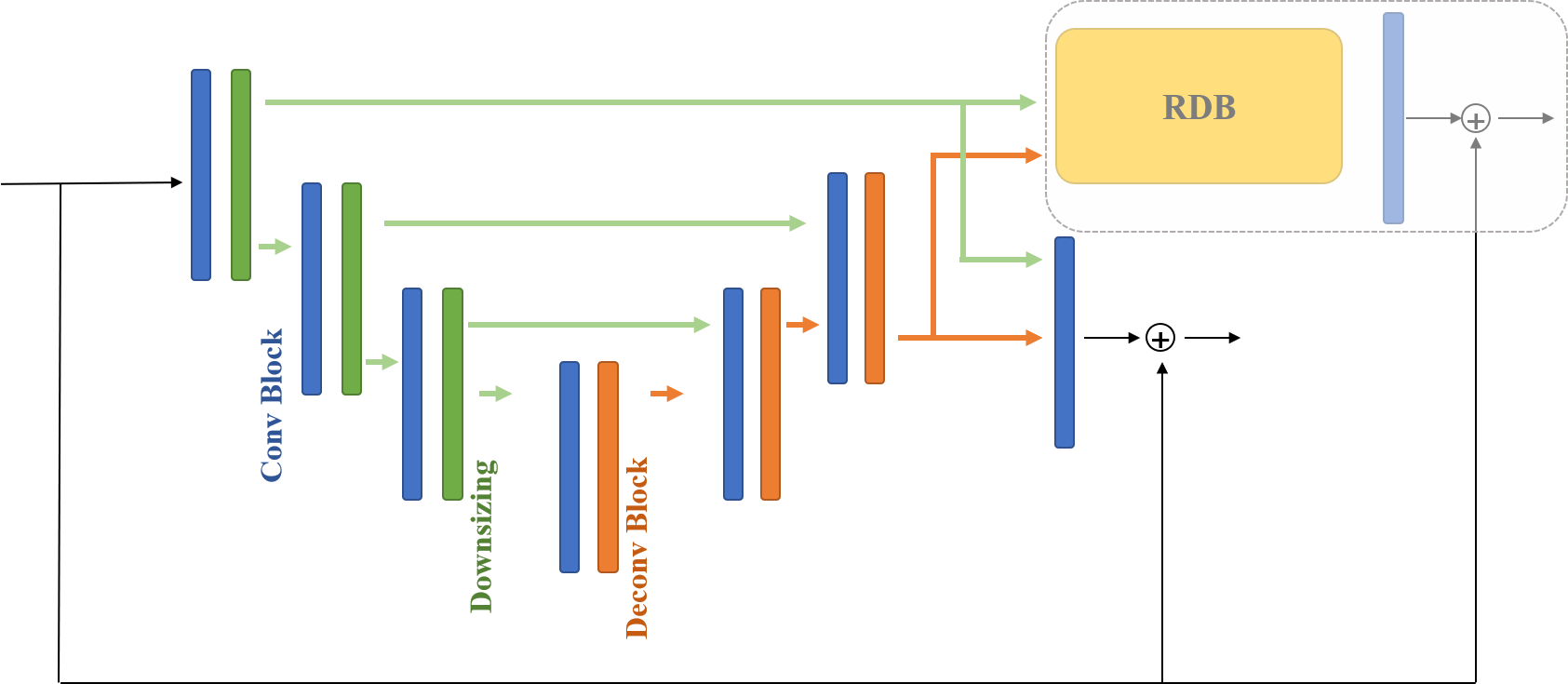}
}
\caption{\small{The model architecture proposed by ENERZAi Research team. Semitransparent residual dense block belongs to the super-network and is detached after training.}}
\label{fig:ENERZAi1}
\end{figure}

The solution proposed by ENERZAi Research is inspired by the \textit{Once-for-All} approach~\cite{cai2019once} and consists of two models: one super-network and one sub-network. They both share the same U-Net~\cite{ronneberger2015u} like module, and the difference comes from their top layers: the sub-network has one convolutional layers, while the super-network additionally contains several residual dense blocks as shown in Fig.~\ref{fig:ENERZAi1}. Both models are first trained jointly using a combination of $L_1$ and \textit{MS-SSIM} loss functions. The super-network is then detached after the PSNR score goes above a predefined  threshold, and the sub-net is further fine-tuned alone. The model was trained on 256$\times$256px patches using Adam optimizer with a batch size of 8 and a learning rate of $1e-3$. The resulting model is able to process 2432$\times$3000px images under 300 ms on the Samsung Galaxy S21 smartphone.

\begin{figure}[h!]
\centering
\resizebox{0.76\linewidth}{!}
{
\includegraphics[width=1.0\linewidth]{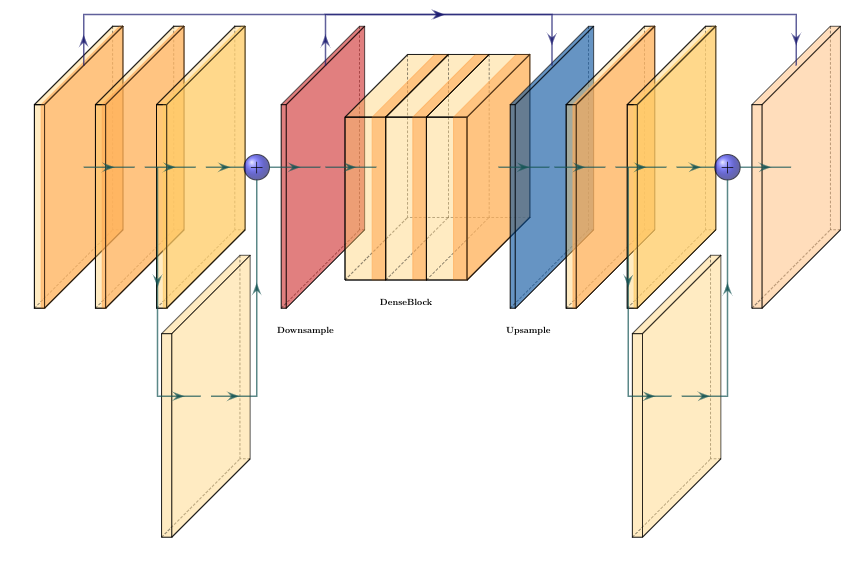}
} \resizebox{0.7\linewidth}{!}
{
\includegraphics[width=1.0\linewidth]{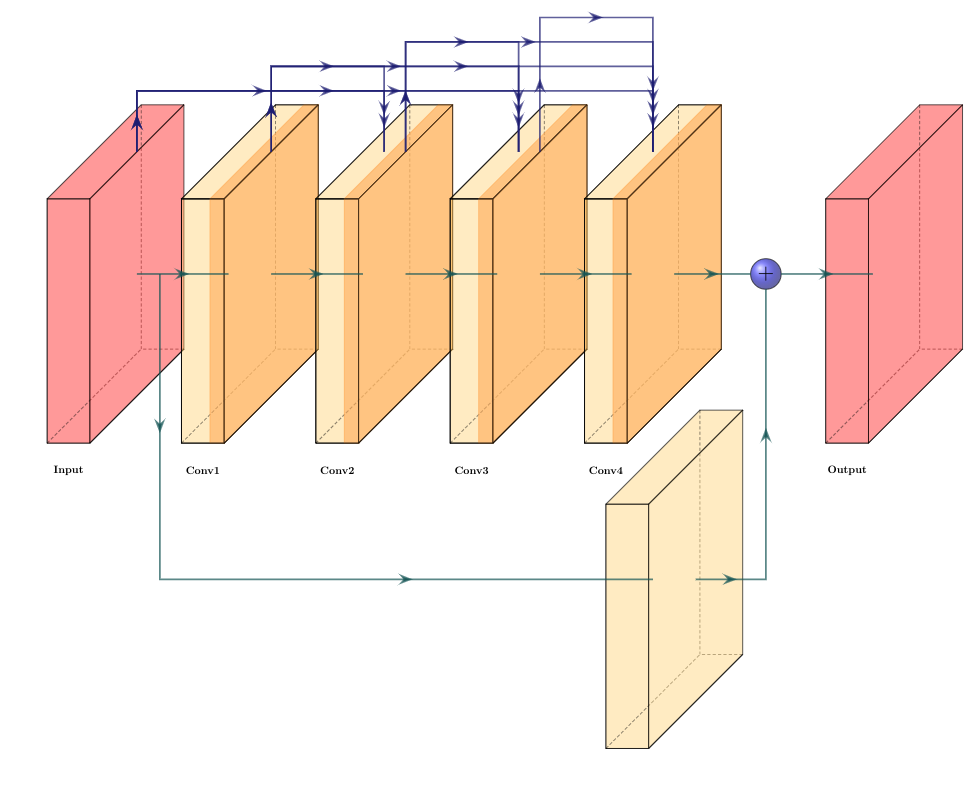}
}
\caption{\small{A shallow U-Net based model (top) and a densely connected block (bottom) proposed by ENERZAi Research team.}}
\label{fig:ENERZAi2}
\end{figure}

The second model proposed by this team (which did not officially participate in the final test phase) is demonstrated in Fig.~\ref{fig:ENERZAi2}. Same as above, the authors started from a standard U-Net based architecture and inserted an additional dense block with skip connections~\cite{huang2017densely} in its bottleneck layer. The authors used \textit{PReLU} activations to get better fidelity results and compressed the model using knowledge distillation technique~\cite{hinton2015distilling,heo2019comprehensive}. The model was trained with a combination of $L_1$ and \textit{MS-SSIM} losses using a dynamic learning rate~\cite{smith2017cyclical} ranging from $1e-4$ to $5e-7$.

\subsection{MOMA-Denoise}

\begin{figure}[h!]
\centering
\resizebox{0.9\linewidth}{!}
{
\includegraphics[width=1.0\linewidth]{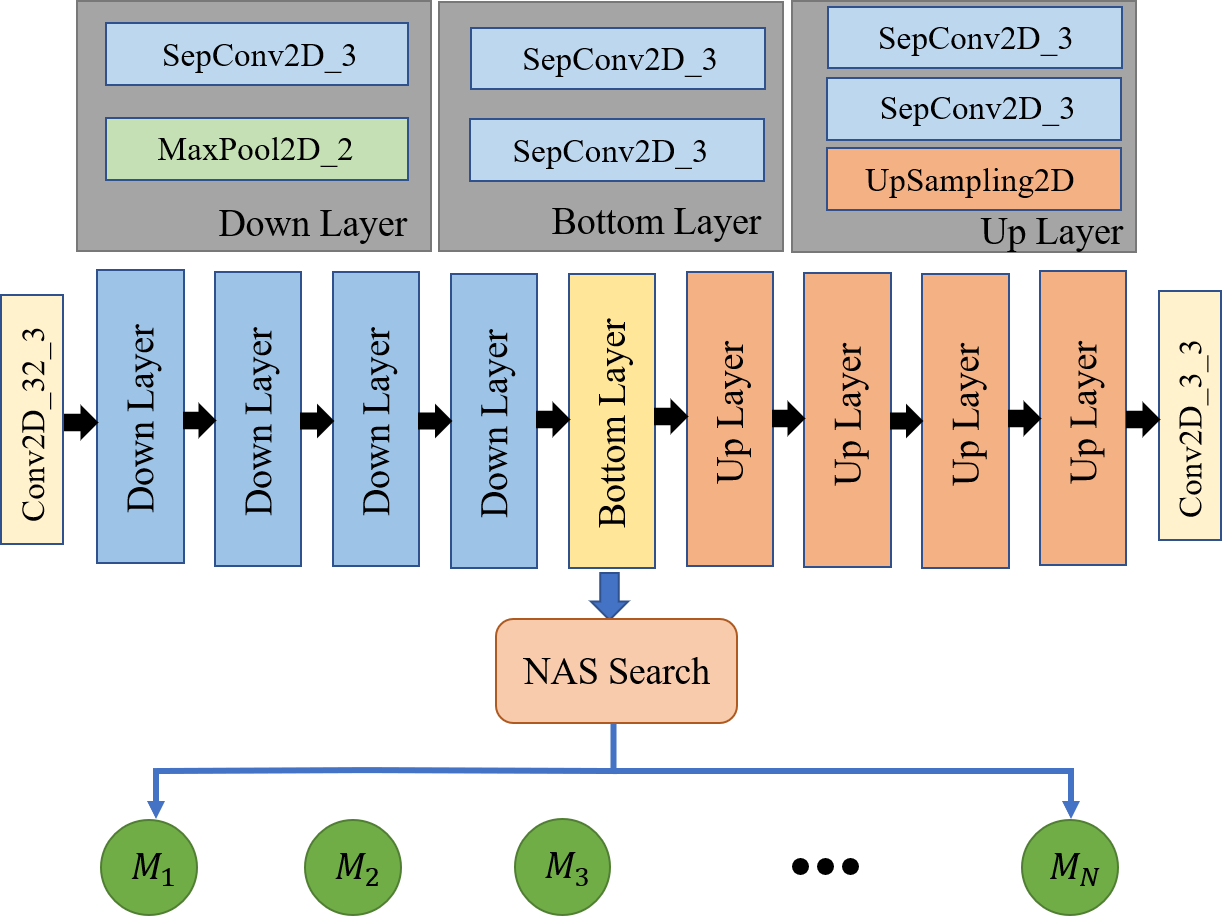}
}
\caption{\small{The training NAS-based pipeline and the architecture proposed by MOMA-Denoise team.}}
\label{fig:MOMA}
\end{figure}

MOMA-Denoise team used an in-house Xiaomi AutoML Framework-MOMA to find the best architecture with Neural Architcture Search. The authors used a light-weight U-Net based network consisting of separable convolutions, maxpooling and upsampling layers as a base model, and searched for the best filter and channel sizes using NAS (Fig.~\ref{fig:MOMA}). To get more training data, the authors unprocessed the provided JPEG images to RAW format, added artificial Poisson-Gaussian noise and mapped the resulting images back to RGB format to mimic real noise present on smartphone photos. The considered image processing pipeline included black level correction, digital gain, demosaicing, device RGB to sRGB mapping, gamma correction and global tone mapping operations. The model was trained to minimize $L_1$ loss using Adam optimizer with a learning rate of $1e-3$ decreased to $1e-7$ within the training process.

\subsection{Mier}

\begin{figure}[h!]
\centering
\resizebox{0.9\linewidth}{!}
{
\includegraphics[width=1.0\linewidth]{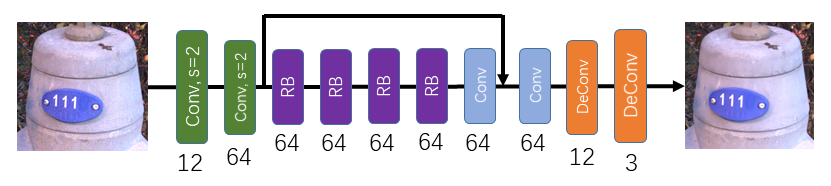}
}
\caption{\small{A small U-Net model proposed by Mier team.}}
\label{fig:Mier}
\end{figure}

Team Mier developed a small U-Net like architecture presented in Fig.~\ref{fig:Mier}. This model starts with two convolutions with a stride of 2 to reduce the size of the features, followed by four residual blocks with skip connections and two deconvolutional layers. During the training, the authors used asymmetric convolutions~\cite{liu2020mmdm,ding2019acnet} to enhance the kernel skeleton, which were fused after training. The model was trained on 256$\times$256px patches, \textit{Charbonnier} loss was used as a target fidelity metric. The model parameters were optimized using Adam with a batch size of 16 and an initial learning rate of $1e-4$ decreased to $1e-7$ using the cosine annealing strategy. It should be mentioned that the original model was trained in PyTorch and then exported to TensorFlow and TFLite formats.

\subsection{GdAlgo}

\begin{figure}[h!]
\centering
\resizebox{0.9\linewidth}{!}
{
\includegraphics[width=1.0\linewidth]{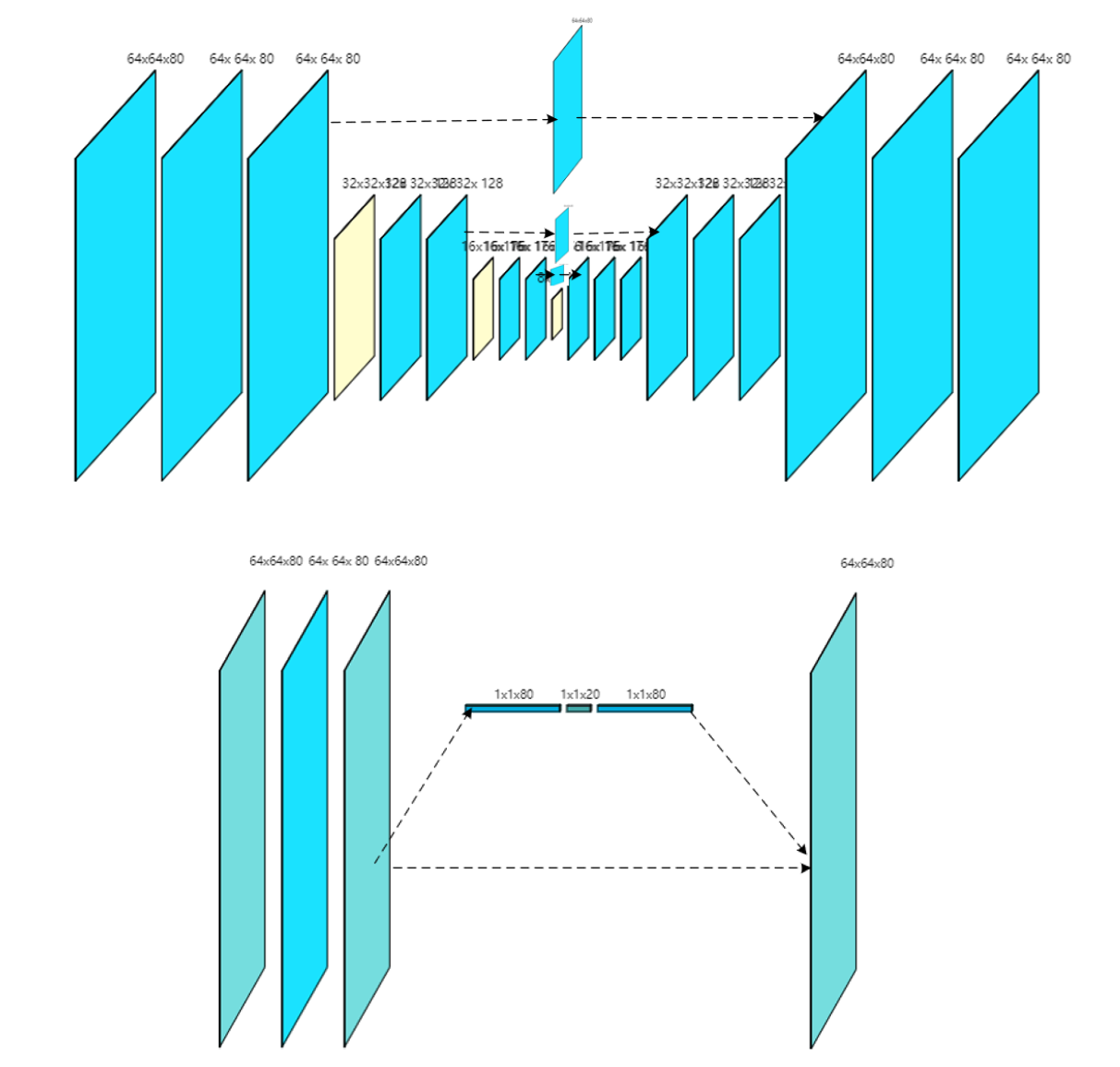}
}
\caption{\small{Guided Attention U-Net and a channel attention block designed by GdAlgo team.}}
\label{fig:GdAlgo}
\end{figure}

Team GdAlgo proposed a Guided Attention U-Net (GAU-net) model demonstrated in Fig.~\ref{fig:GdAlgo}. The standard convolutional layers in this model are replaced by channel attention blocks to get better visual and fidelity results. Inspired by the Multi-Stage Image Restoration approach~\cite{zamir2021multi}, the authors adopted the corresponding multi-stage learning strategy. The model was trained to minimize \textit{Charbonnier} loss function, the edge loss was discarded as no improvement was observed in the experiments.

\section{Additional Literature}

An overview of the past challenges on mobile-related tasks together with the proposed solutions can be found in the following papers:

\begin{itemize}
\setlength\itemsep{0.2mm}
\item Image Denoising:\, \cite{abdelhamed2020ntire,abdelhamed2019ntire}
\item Learned End-to-End ISP:\, \cite{ignatov2019aim,ignatov2020aim}
\item Perceptual Image Enhancement:\, \cite{ignatov2018pirm,ignatov2019ntire}
\item Bokeh Effect Rendering:\, \cite{ignatov2019aimBokeh,ignatov2020aimBokeh}
\item Image Super-Resolution:\, \cite{ignatov2018pirm,lugmayr2020ntire,cai2019ntire,timofte2018ntire}
\end{itemize}

\section*{Acknowledgements}

We thank Samsung Electronics, AI Witchlabs and ETH Zurich (Computer Vision Lab), the organizers and sponsors of this Mobile AI 2021 challenge.

\appendix
\section{Teams and Affiliations}
\label{sec:apd:team}

\bigskip

\subsection*{Mobile AI 2021 Team}
\noindent\textit{\textbf{Title: }}\\ Mobile AI 2021 Real Image Denoising Challenge\\
\noindent\textit{\textbf{Members:}}\\ Andrey Ignatov$^{1,3}$ \textit{(andrey@vision.ee.ethz.ch)}, Kim Byeoung-su$^2$ \textit{(rui.kim@samsung.com)}, Radu Timofte$^{1,3}$  \textit{(radu.timofte@vision.ee.ethz.ch)}\\
\noindent\textit{\textbf{Affiliations: }}\\
$^1$ Computer Vision Lab, ETH Zurich, Switzerland\\
$^2$ Samsung Electronics, South Korea\\
$^3$ AI Witchlabs, Switzerland\\

\subsection*{NOAHTCV}
\noindent\textit{\textbf{Title:}}\\Efficient and Specialized Network Search for Image Denoising\\
\noindent\textit{\textbf{Members:}}\\ \textit{Fenglong Song (songfenglong@huawei.com)}, Cheng Li, Shuai Xiao, Zhongqian Fu, Matteo Maggioni, Yibin Huang\\
\noindent\textit{\textbf{Affiliations: }}\\
Huawei Noah's Ark Lab, China\\
\url{http://www.noahlab.com.hk/}\\

\subsection*{MegDenoise}
\noindent\textit{\textbf{Title:}}\\Fast Image Denoise network with Bottleneck Encoder and Slight Decoder\\
\noindent\textit{\textbf{Members:}}\\ \textit{Shen Cheng (chengshen@megvii.com)}, Xin Lu, Yifeng Zhou, Liangyu Chen, Donghao Liu, Xiangyu Zhang, Hao-
qiang Fan, Jian Sun, Shuaicheng Liu\\
\noindent\textit{\textbf{Affiliations: }}\\
Megvii, China\\

\subsection*{ENERZAi Research}
\noindent\textit{\textbf{Title:}}\\Learning Small Denoising U-Net by Shrinking Large Network\\
\noindent\textit{\textbf{Members:}}\\ \textit{Minsu Kwon (minsu.kwon@enerzai.com)}, Myungje Lee, Jaeyoon Yoo, Changbeom Kang, Shinjo Wang\\
\noindent\textit{\textbf{Affiliations: }}\\
ENERZAi, Seoul, Korea \\ \textit{enerzai.com}\\

\subsection*{MOMA-Denoise}
\noindent\textit{\textbf{Title:}}\\UnetS: A Lightweight Denoise Model\\
\noindent\textit{\textbf{Members:}}\\ \textit{Bin Huang (2659934122@qq.com)}, Tianbao Zhou\\
\noindent\textit{\textbf{Affiliations: }}\\
Xiaomi AI-Lab, China \\

\subsection*{Mier}
\noindent\textit{\textbf{Title:}}\\Small UNet\\
\noindent\textit{\textbf{Members:}}\\ \textit{Shuai Liu$^{1}$ (18601200232@163.com)}, Lei Lei$^{2}$, Chaoyu Feng$^{2}$\\
\noindent\textit{\textbf{Affiliations: }}\\
$^1$ North China University of Technology, China \\
$^2$ Xiaomi Inc., China \\

\subsection*{GdAlgo}
\noindent\textit{\textbf{Title:}}\\Guided Attention U-Net for Image Restoration\\
\noindent\textit{\textbf{Members:}}\\ \textit{Liguang Huang (huang.liguang@qq.com)}, Zhikun Lei, Feifei Chen\\
\noindent\textit{\textbf{Affiliations: }}\\
Algorithm Department, Goodix Technology, China \\

{\small
\bibliographystyle{ieee_fullname}

}

\end{document}